\documentclass[10pt,secnumarabic,twocolumn,nofootinbib,showkeys,showpacs,]{revtex4}
\usepackage{graphicx}
\usepackage{epsf}
\usepackage{bm}
\usepackage{amsmath}
\usepackage{amsfonts}
\usepackage{amssymb}
\usepackage{epstopdf}
\usepackage{color}
\begin{document}
\title{A new interacting two fluid model and its consequences}
\author{German S. Sharov}
\email{german.sharov@mail.ru}
\affiliation{Department of Mathematics, Tver State University, 170002, Sadovyj per. 35, Tver, Russia}
\author{Subhra Bhattacharya}
\email{subhra.maths@presiuniv.ac.in}
\affiliation{Department of Mathematics, Presidency University, Kolkata-700073, West Bengal, India}
\author{Supriya Pan}
\email{span@iiserkol.ac.in}
\affiliation{Department of Physical Sciences, Indian Institute of Science Education and Research$-$Kolkata, Mohanpur$-$741246, West Bengal, India}
\author{Rafael C. Nunes}
\email{rcnunes@fisica.ufjf.br}
\affiliation{Departamento de F\'{\i}sica, Universidade Federal de Juiz de Fora, 36036-330, Juiz de Fora, MG, Brazil}
\author{Subenoy Chakraborty}
\email{schakraborty@math.jdvu.ac.in}
\affiliation{Department of Mathematics, Jadavpur University, Kolkata-700032, West Bengal, India}
%%%%%%%%%%%%%%%%%%%%%%%%%%%%%%%%%%%%%%%%%%%%%%%%%%%%%%%%%%%%%%%%%%%%%%%%%
%\documentclass[a4paper,fleqn,usenatbib]{mnras}

%\usepackage[T1]{fontenc}
%\usepackage{ae,aecompl}

%%%%% AUTHORS - PLACE YOUR OWN PACKAGES HERE %%%%%

% Only include extra packages if you really need them. Common packages are:
%\usepackage{graphicx}   % Including figure files
%\usepackage{amsmath}    % Advanced maths commands
%\usepackage{amssymb}    % Extra maths symbols
%\usepackage{color}
%\newtheorem{remark}{Remark}[section]
%\usepackage{booktabs}
%\usepackage{multirow}
%\usepackage{siunitx}
%\usepackage{rotating}
%\usepackage{pdflscape}
%\usepackage{adjustbox}
%%%%%%%%%%%%%%%%%%%%%%%%%%%%%%%%%%%%%%%%%%%%%%%%%%%%%%%%%%%%%%%%%%%%%%%%%%%%%%%%%%%%
%%%%%%%%%%%%%%%%%%%%%%%%%%%%%%%%%%%%%%%%%%%%%%%%%%%%%%%%%%%%%%%%%%%%%%%%%%%%%%%%
%%%%%%%%%%%%%%%%%%%%%%%%%%%%%%%%%%%%%%%%%%%%%%%%%%%%%%%%%%%%%%%%%%%%%%%%%%%%%%%%%%%%%%%%%%%%%%%%%%%%%%%%%%%%%%%%%%
%%%%%%%%%%%%%%%%%%%%%%%%%%%%%%%%%%%%%%%%%%%%%%%%%%%%%%%%%%%%%%%%%%%%%%%%%%%%%%%%%%%%%%%%%%%%%%%%%%%%%%%%%%%%%%%%%%
\begin{abstract}

In the background of a homogeneous and isotropic spacetime with zero spatial curvature,
we consider interacting scenarios
between two barotropic fluids, one is the pressureless  dark matter (DM) and the other one is dark energy (DE), in which the equation of state (EoS) in DE is either constant or time dependent. In particular, for constant EoS in DE, we show that the evolution equations for both fluids can be analytically solved. For all these scenarios, the model parameters have been constrained using the current astronomical observations from Type Ia Supernovae, Hubble parameter measurements, and baryon acoustic oscillations distance measurements. Our analysis shows that both for constant and variable EoS in DE, a very small but nonzero interaction in the dark sector is favored while the EoS in DE can predict a slight phantom nature, i.e. the EoS in DE can cross the phantom divide line `$-1$'.
On the other hand, although the models with variable EoS describe the observations better, but the Akaike Information Criterion supports models with minimal number of parameters. However, it is found that all the models are very close to the $\Lambda$CDM cosmology. 

\end{abstract}

%%%%%%%%%%%%%%%%%%%%%%%%%%%%%%%%%%%%%%%%%%%%%%%%%%%%%%%%%%%%%%%%%%%%%%%%%%%%%%%%%%%%%%%%%%%%%%%%%%%%%%%%%%%%%%
\maketitle
%%%%%%%%%%%%%%%%%%%%%%%%%%%%%%%%%%%%%%%%%%%%%%%%%%%%%%%%%%%%%%%%%%%%%%%%%%%%%%%%%%%%%%%%%%%%%%%%%%%%%%%%%%%%%%%%%%
%~~~\myclassification{98.80.Cq, 98.80.-k}\\\\
%%%%%%%%%%%%%%%%%%%%%%%%%%%%%%%%%%%%%%%%%%%%%%%%%%%%%%%%%%%%%%%%%%%%%%%%%%%%%%%%%%%%%%%%%%%%%%%%%%%%%%%%%%%%%%%%%%
\section{Introduction}
\label{Intro}

The explanation of the late-time accelerated expansion of the universe \cite{Perlmutter1999, Riess1998, Bernardis2000,  Percival2001, Spergel2003, JT2003, Tegmark2004, Eisenstein2005, Komatsu2011} has become one of the biggest and open problems in modern cosmology today. The most reasonable description for this accelerating phase introduces some hypothetical dark energy (DE) fluid comprising about 70\% of the total energy density of the universe \cite{Planck2015}. The cosmological constant, $\Lambda$, associated with the zero point energy of the quantum fields, is the main candidate for DE in agreement with a large number of available observational data. However, despite of great success of  $\Lambda$-cosmology, it presents serious objections in the interface of cosmology and particle physics, such as the cosmological constant problem \cite{Weinberg1989, Padmanabhan2003} and the coincidence problem \cite{Steinhardt2003}. Due to these two serous problems in $\Lambda$, several alternatives have been proposed and discussed in the last several years \cite{AT2010}.

However, concerning different cosmological theories, the scenario where dark matter (DM) interacts with dark energy (DE) has gained much attention in the current literature. The interaction between DM and DE were primarily motivated to address the small value of the cosmological constant, however later on  the models were found to provide a reasonable explanation to the cosmic coincidence problem \cite{Bolotin2015, Wang2016}. Recently it has been argued that the current observational data can favor the late-time interaction between DM and DE \cite{Salvatelli2014, SVP2015, SPVN2016, NPS2016, KN2016, Bruck2016}. The coupling parameter of the interaction in the dark sector has also been measured by several observational data \cite{YGC2015, NPS2016, KN2016, Bruck2016, Weiqiang1, Weiqiang2, Weiqiang3, Xia2016, CT2016, Murgia2016}. In fact, the class of interacting DM $-$ DE models could be a promising candidate to resolve the current tensions on $\sigma_8$ and the local value of the Hubble constant $H_0$ \cite{KN2016, PTram2016}. Further, it has been contended that the interaction between these dark sectors can influence on the perturbation analysis which results in significant changes in the lowest multipoles of the CMB spectrum \cite{Zimdahl2005, WZLAM2007}.
For a comprehensive analysis on several interacting models explored in the last couple of years we refer  \cite{Amendola2000, BC2000, ZPC2001, Dalal2001, HPZ2004, AQ2003, MMP2003, PZ2005, CJPZ2003, He2008, Quartin2008, Bohmer2008, CMU2009, NO2010, Hashim2014, PBC2015, Nicola2015, Skordis2015, Marra2016, Ankan}. Therefore, based on the analysis on interacting dark energy models with a special indication for a nonzero interaction between the dark sectors from the the currently available observational data, it seems promising that the interaction in the dark sectors might open some new possibilities in near future.

The interacting dynamics proposes for continuous energy and/or momentum exchange between DM and DE with
the evolution of the universe and  modifies the conservation equations for DM ($\equiv$ $T_{\mu\nu}^{DM}$) and DE ($\equiv$ $T_{\mu\nu}^{DE}$) as
\begin{eqnarray}\label{eqn1}
\nabla^{\nu} T_{\mu\nu}^{DM}&=& Q,~~~\mbox{and}~~\nabla^{\nu} T_{\mu\nu}^{DE}= -Q,
\end{eqnarray}
%%%%%%%%%%%%%%%%%%%%%%%%%%%%%%%%%%%%%%%%%%%%%%%%%%%%%%%%%%%%%%%%%%%%%%%%%%%%%%%%%%%%%%%%%%%%%%%%%%%%%%%%%%%%%%%%%%%%%%%%%%%%%%%%%%%%%%%%%%%%%%%%%%%%%%%
where the function $Q$ characterizes the interaction between the dark sectors in which $Q> 0$ represents the energy and/or momentum flow from DE to DM while $Q < 0$ means energy and/or momentum flow takes place from DM to DE. Thus, essentially, we choose different functional forms for $Q$ (with correct physical dimension) and we solve the corresponding evolution equations. The procedure followed is similar to that when one chooses different parametrizations for the EOS in DE. Few developments can be found in the literature where the complete dynamics of the interacting model has been explored \cite{PBC2015, Chimento2010, Chimento2012}. Another approach to consider the interaction between two fluids is for instance the sigma models \cite{Chervon2013, PT2014}.

In the present work, we determine the complete dynamics of an interacting scenario where DE interacts with pressureless DM through a nongravitational interaction considered in \cite{Sanchez2014} and we generalize its observational consequences considering that the DE component is barotropic and its EoS could be either constant or time dependent. We first consider the constant equation of state in dark energy and show that the background evolution can be analytically solved which we constrain using the combined analysis from Type Ia Supernovae, Hubble data, and baryon acoustic oscillation data. Then we apply the same combined analysis of the above observational data to the case when the EoS in DE is evolving with the cosmic time.

The work has been organized in the following way: Section \ref{ide-flrw} presents a brief
introduction of the interacting dynamics in a spatially flat Friedmann-Lema\^{i}tre-Robertson-Walker (FLRW) universe. In section \ref{sec-analytic},
we present the analytic solutions for dark energy and cold dark matter for constant equation
of state $\omega_d$ $\neq -1$ and when the dark energy is the cosmological constant while
additionally, in subsection \ref{sec:asymptotic}, we perform an asymptotic analysis of the
analytic solutions, and then we discussed about the possible sign of the interaction. In section \ref{sec-variable-de} we introduce dynamical DE models interacting with CDM.
Section \ref{data-analysis} contains all the data analysis tools we used in this study. In the next section \ref{results} we present the observational constraints on the interacting models and the main results of the paper. Finally, we close our work in section \ref{discu} with summary and discussions.

\section{Interacting Dynamics in  flat FLRW}
\label{ide-flrw}

Observations from cosmic microwave background radiation and large scale structure predict in good approximation that our Universe is homogeneous and isotropic in the largest scale, and further it is almost spatially flat \cite{Planck2015}. This line element describing such a universe is known as the Friedmann-Lema\^{i}tre-Robertson-Walker line element, which takes the form
\begin{eqnarray}
ds^2= -dt^2+ a^2 (t) (dx^2+ dy^2+ dz^2),\label{FLRW-metric}
\end{eqnarray}
%%%%%%%%%%%%%%%%%%%%%%%%%%%%%%%%%%%%%%%%%%%%%%%%%%%%%%%%%%%%%%%%%%%%%%%%%
where $a(t)$ is the scale factor of the universe, $t$
being the universal cosmic time. Further, this homogeneous and isotropic principle
gives a restriction on the matter distribution of our Universe. It tells that the
matter distribution should be of perfect fluid type which takes its
energy momentum tensor as $T_{\mu \nu}= (p+\rho) u_{\mu} u_{\nu}+ p g_{\mu \nu}$, where
$\rho$, $p$, respectively denote the energy density and the pressure of
the cosmic fluid, and $u_{\nu}$ is the four velocity vector of the fluid components.
For our model, we consider the general relativity described by
the Einstein's field equations $G_{\mu \nu}= 8 \pi G T_{\mu \nu}$, and two fluids as follows:\\

$\bullet$ Pressureless dark matter (or dust) where $\rho_m$ is its energy density and $p_m (= 0)$, is its pressure.\\

$\bullet$ A dark energy fluid satisfying a barotropic equation of state $p_d= \omega_d \rho_d$, where $\rho_d$, $p_d$,
are respectively the energy density and the pressure of the component, $\omega_d$ is its equation of state. \\

Thus, for a co-moving observer ($u^{\mu}= \delta^{\mu}_0$), the Einstein's field equations are explicitly written as (considering $8 \pi G= 1$)
\begin{align}
H^2 &= \frac{1}{3}(\rho_m+ \rho_d),\label{friedmann1}\\
\dot{H} &= -\frac{1}{2} (p_m+ p_d+ \rho_m+\rho_d),\label{friedmann2}
\end{align}
%%%%%%%%%%%%%%%%%%%%%%%%%%%%%%%%%%%%%%%%%%%%%%%%%%%%%%%%%%%%%%%%%%%%%%%%%%%%%%%%%%%%%%%%%%%%%%
where $H= \dot{a}/a$, is the Hubble parameter.
%%%%%%%%%%%%%%%%%%%%%%%%%%%%%%%%%%%%%%%%%%%%%%%%%%%%%%%%%%%%%%%%%%%
Further, the conservation equations for DM and DE follows from eqn. (\ref{eqn1})
\begin{eqnarray}
\dot{\rho}_m+3 H \rho_m&=& Q,\label{conservation1}\\
\dot{\rho}_d+3 H (1+\omega_d) \rho_d&=& -Q\, .\label{conservation2}
\end{eqnarray}
%%%%%%%%%%%%%%%%%%%%%%%%%%%%%%%%%%%%%%%%%%%%%%%%%%%%%%%%%%%%%%%%%%%%%%%%%%%%%%
Clearly, equations (\ref{conservation1}) and (\ref{conservation2}) offer a new dynamics of the universe with the use of the interaction function $Q$. 
There are many proposed interactions in the literature to study
the dynamics of the universe, however, the exact functional form of $Q$ is still unknown. From the coupled continuity equations (\ref{conservation1})
and (\ref{conservation2}), we observe that $Q$ could be any arbitrary function
of the parameters $H$, $\rho_m$, $\rho_d$. So, naturally, one can construct infinitely many interacting models to understand the dynamics of the universe in this framework.
In the present work, we start with the following interaction \cite{Sanchez2014}:
\begin{eqnarray}
Q&=&  \alpha H (\rho_m^\prime+\rho_d ^\prime).\label{interaction}
\end{eqnarray}
%%%%%%%%%%%%%%%%%%%%%%%%%%%%%%%%%%%%%%%%%%%%%%%%%%%%%%%%%%%%%%%%%%%%%%%%%%%%%%%%%%%%%%%%%
where $\alpha$ is the coupling parameter which is dimensionless, and the `~$^\prime$~' denotes
the differentiation with respect to $x= \ln \left(a/a_0 \right)$ ($a_0$ is the present value of
the scale factor which has been set to be unity in this work, i.e. $a_0= 1$). In fact, if one uses the conservation equations
(\ref{conservation1}), and (\ref{conservation2}), it will be readily clear that the interaction (\ref{interaction}) has
the following equivalent form
\begin{eqnarray}\label{eq-int}
Q &=& - 3\, \alpha H (\rho_m+ \rho_d+ \omega_d\, \rho_d)
\end{eqnarray}
which directly includes the equation of state parameter of the dark energy component, that means the
pressure of the dark energy component.
However, inserting (\ref{interaction}) into the conservation equations (\ref{conservation1})
and (\ref{conservation2}), we decouple the conservation equations as
\begin{eqnarray}
(1-\alpha) \rho_m^ \prime + 3 \rho_m- \alpha \rho_d ^ \prime&=& 0,\label{conservation1.1}\\
(1+ \alpha) \rho_d ^ \prime + + 3 \rho_d (1+ \omega_d)+ \alpha \rho_m ^ \prime &=& 0 ,\label{conservation2.1}
\end{eqnarray}
%%%%%%%%%%%%%%%%%%%%%%%%%%%%%%%%%%%%%%%%%%%%%%%%%%%%%%%%%%%%%%%%%%%%%%%%%%%%%%%%%%%%%%%%%%%%%%%%%%%
where both equations (\ref{conservation1.1}) and (\ref{conservation2.1}) seem to describe an
effective non-interacting two fluid system, and due to the interaction, the effective
EoS for pressureless DM ($\omega^{eff}_m$), DE ($\omega^{eff}_d$) take the forms
\begin{eqnarray}
\omega^{eff}_m&=& -\frac{Q}{3 H \rho_m}\nonumber\\
&=& -\frac{\alpha}{\rho_m} (\rho^\prime_m+ \rho^\prime_d)\nonumber\\
&=& \frac{3\,\alpha}{r} (1+r+ \omega_d),\label{CDM-eff}
\end{eqnarray}
%%%%%%%%%%%%%%%%%%%%%%%%%%%%%%%%%%%%%%%%%%%%%%%%%%%%%%%
and
\begin{eqnarray}
\omega^{eff}_d&=& \omega_d+ \frac{Q}{3 H \rho_d},\nonumber\\
&=& \omega_d+ \frac{\alpha}{\rho_d} (\rho^\prime_m+ \rho^\prime_d),\nonumber\\
&=& \omega_d - 3\,\alpha \,(1+r+ \omega_d),\label{DE-eff}
\end{eqnarray}
%%%%%%%%%%%%%%%%%%%%%%%%%%%%%%%%%%%%%%%%%%%%%%%%%%%%%%%%%%%%%%%
where $r= \rho_m/\rho_d$, is the coincidence parameter.
Now, from (\ref{CDM-eff}), (\ref{DE-eff}), we note that, even if the EoS in DM and DE
 are constant, but due to the interaction between them, the effective EoS of the two components become variable in nature.
Now, if we denote the total EoS for the non-interacting DM-DE system by $\omega^{eff}= \omega_m+ \omega_d= \omega_d$ (as DM is pressureless, i.e. $\omega_m= 0$), then the effective EoS: $\omega^{eff}_t= w^{eff}_m+ \omega^{eff}_d$, of the same system due to this interaction becomes
\begin{eqnarray}
\omega^{eff}_t&=& \omega^{eff}+ \alpha \left(\frac{1}{\rho_d}-\frac{1}{\rho_m}\right) (\rho^\prime_m+ \rho^\prime_d)\nonumber\\
&=& \omega_d+ \alpha \left(\frac{1}{\rho_d}-\frac{1}{\rho_m}\right) (\rho^\prime_m+ \rho^\prime_d),\nonumber\\
&=& \omega_d + \frac{3\,\alpha}{r} (1-r) (1+r+\omega_d)\label{total-eff}
\end{eqnarray}
%%%%%%%%%%%%%%%%%%%%%%%%%%%%%%%%%%%%%%%%%%%%%%%%%%%%%%%%%%%%%%%%%%%%%%%%%%%%%%%%%%%%%%%%%%%%%%%%%%%%%%%%%%%%%%%%%%%%%%%%%%%

Equations (\ref{CDM-eff}), (\ref{DE-eff}) lead to the following interesting possibilities we will discuss now. \\

$\bullet$ If $Q> 0$, Eq. (\ref{CDM-eff}) implies that, $\omega_m^{eff}< 0$ meaning that the effective equation of state of the pressureless dark matter\footnote{We define the effective EoS of the pressureless dark matter as the EoS of the pressureless dark matter plus the energy gained due to this interacting process.} is of exotic type, and it could become a dark energy candidate (EoS $< -1/3$) if $Q> H \rho_m$. On the other hand, the effective equation of state for dark energy satisfies the following inequality: $\omega_d< \omega_d^{eff}< \infty$, which shows that if the EoS in dark energy satisfies the relation $\omega_s \leq -1$, then the effective interactive system is of course in the quintessence era ($\omega_d^{eff}> -1$). However, if the dark energy is of phantom kind, then the above inequality could still hold, but in this case, the effective EoS in dark energy may lie in the phantom/quintessence region depending on the strength of the interaction.\newline

$\bullet$ If $Q< 0$, then $\omega_m^{eff}> 0$ irrespective of anything. In case of effective dark energy, we find that $\omega_d^{eff}< \omega_d$ which readily points that, for $\omega_d \leq -1$, the effective equation of state is always phantom in nature. On the other hand, for $\omega_d> -1$, $\omega_d^{eff}$ could cross the phantom barrier line `$-1$' due to the presence of $Q/ 3 H \rho_d< 0$ in (\ref{DE-eff}). But, the possibility of quintessence behavior can't be excluded as again we do not specify the strength of the interaction.\newline

However, introducing the total energy
density of the universe, $\rho_t= \rho_m+ \rho_d$, we can express the dark energy
and the dark matter density in terms of the total energy density and its derivative as
\begin{eqnarray}
\rho_d =-\left(\frac{\rho^\prime _t+ 3 \rho_t}{3 \omega_d}\right),\label{DE-density}\\
\rho_m=\left(\frac{\rho^\prime _t+ 3 (1+ \omega_d)\rho_t}{3 \omega_d}\right),\label{DM-density}
\end{eqnarray}
%%%%%%%%%%%%%%%%%%%%%%%%%%%%%%%%%%%%%%%%%%%%%%%%%%%%%%%%%%%%%%%%%%%%%%%%%%%%%%%%%%%%%%%%%%%%%%%%%%
Now, inserting (\ref{DE-density}) into (\ref{conservation2.1}), we obtain the following second order
differential equation
\begin{align}
\rho^{\prime\prime}_t+ 3 \left[2+ \omega_d- \alpha \omega_d- \frac{\omega^\prime_d}{3\omega_d}\right]\rho^\prime_t+ 9 \left[(1+\omega_d)-\frac{\omega^\prime_d}{3\omega_d}\right]\rho_t & =  0,\label{ode1}
\end{align}
%%%%%%%%%%%%%%%%%%%%%%%%%%%%%%%%%%%%%%%%%%%%%%%%%%%%%%%%%%%%%%%%%%%%%%%%%%%%%%%%%%%%%%%%%%%%
which if we could solve might describe the possible dynamics of the universe. In what follows, we consider the evolution of the universe both for constant and dynamic EoS in DE.

\section{Constant EoS in DE}
\label{sec-analytic}

Let us now consider the case when the EoS in DE is independent of time, we denote the EoS by $\omega_{d}$. This consideration reduces the differential equation (\ref{ode1}) into the following simplified format
\begin{eqnarray}
\rho^{\prime\prime}_t+ 3 \big[2+ \omega_{d}- \alpha \omega_{d} \big]\rho^\prime_t+ 9 (1+\omega_{d})\rho_t&=& 0,\label{ode2}
\end{eqnarray}
%%%%%%%%%%%%%%%%%%%%%%%%%%%%%%%%%%%%%%%%%%%%%%%%%%%%%%%%%%%%%%%%%%%%%%%%%%%%%%%%%%%%
which solves for $\rho_t$ as
\begin{eqnarray}
\rho_t&=& \rho_1 e^{m_1 x}+ \rho_2 e^{m_2 x}= \rho_1 a^{m_1}+ \rho_2 a^{m_2},\label{solution-constant-EoS}
\end{eqnarray}
%%%%%%%%%%%%%%%%%%%%%%%%%%%%%%%%%%%%%%%%%%%%%%%%%%%%%%%%%%%%%%%%%%%%%%%%%%%%%%%%%%%%
where $\rho_1$, $\rho_2$ are the constants of integration, and, $\big(m_1, m_2\big)$
are given by
\begin{eqnarray}
m_1&=& \frac{3}{2}\left[-(2+\omega_{d}-\alpha \omega_{d})+ \sqrt{(1-\alpha)^2 \omega_{d}^2- 4 \alpha \omega_{d}}\right],\label{m1}\nonumber \\
m_2&=& \frac{3}{2}\left[-(2+\omega_{d}-\alpha \omega_{d})- \sqrt{(1-\alpha)^2 \omega_{d}^2- 4 \alpha \omega_{d}}\right],\label{m2}\nonumber
\end{eqnarray}
%%%%%%%%%%%%%%%%%%%%%%%%%%%%%%%%%%%%%%%%%%%%%%%%%%%%%%%%%%%%%%%%%%%%%%%%%%%%%%%%%%%%%%%%%%%%%%%%%%%%%%%%%%%%%%%%%%%%%%%%%%
As $\omega_d< -1/3$, so, $\big(m_1, m_2\big)$ are real. The equation (\ref{solution-constant-EoS}) can
be written as
\begin{eqnarray}
\left(\frac{H}{H_0}\right)^2&=& \Omega_1 (1+z)^{-m_1}+ \Omega_2 (1+z)^{-m_2}.\label{F1}
\end{eqnarray}
%%%%%%%%%%%%%%%%%%%%%%%%%%%%%%%%%%%%%%%%%%%%%%%%%%%%%%%%%%%%%%%%%%%%%%%%%%%%%%%%%%%%%%%%%%%%%%%%%%%%%%%
where $\Omega_1= \rho_1/\rho_{c0}$, $\Omega_2= \rho_2/\rho_{c0}$ in which $\rho_{c0}= 3H_0^2/8\pi G$.
It is interesting to note that, the solution (\ref{F1}) which represents the analytic behavior for two interacting fluids,
has also been found in the context of two non-interacting fluids where one is Brans-Dicke scalar field and the other is the
perfect fluid with constant equation of state \cite{PTBB2016}.

Moreover, using (\ref{DE-density}), (\ref{DM-density}), the explicit analytic expressions for DE and DM density are

\begin{align}
\rho_d& = %-\left(\frac{1}{3 \omega_d}\right)\Bigl[\rho_1(m_1+ 3) (1+z)^{-m_1}+ \rho_2(m_2+ 3) (1+z)^{-m_2}\Bigr]\label{DE-constantA}
 -\frac{\rho_1(m_1+ 3) (1+z)^{-m_1}+ \rho_2(m_2+ 3) (1+z)^{-m_2}}{3 \omega_d}\label{DE-constantA}
\end{align}
%%%%%%%%%%%%%%%%%%%%%%%%%%%%%%%%%%%%%%%%%%%%%%%%%%%%%%%%%%%%%%%%%%%%%
and
%%%%%%%%%%%%%%%%%%%%%%%%%%%%%%%%%%%%%%%%%%%%%%%%%%%%%%%%%%%%%%%%%%%%
\begin{eqnarray}
\rho_m=\left(\frac{1}{3 \omega_{d}}\right)\Bigl[\rho_1(m_1+ 3+ 3\omega_{d}) (1+z)^{-m_1} \nonumber\\+ \rho_2(m_2+ 3+ 3\omega_{d}) (1+z)^{-m_2}\Bigr].\label{DM-constantB}
\end{eqnarray}
%%%%%%%%%%%%%%%%%%%%%%%%%%%%%%%%%%%%%%%%%%%%%%%%%%%%%%%%%%%%%%%%%%%%%%%%%%%%%%%%%%%%%%%%%%%%%%%%%%%
However, the constants $\rho_1$, $\rho_2$ do not serve only as the constants of integrations, they have crucial
meanings. In order to explain this, let us introduce the density parameters for dark matter and dark
energy as $\Omega_m= \rho_m/\rho_c$ ($\rho_c= 3H^2/8\pi G$), $\Omega_d= (\rho_d/\rho_c)$ respectively,
Thus, the Friedmann equation (\ref{friedmann1}) can be written as $\Omega_m+ \Omega_d= 1$. Now, the energy
densities at present time can be given by
\begin{align}
\rho_{d0} &=  -\left(\frac{1}{3 \omega_{d0}}\right)\Bigl[\rho_1(m_1+ 3)+ \rho_2(m_2+ 3)\Bigr],\label{DE-present}\\
\rho_{m0} &= \left(\frac{1}{3 \omega_{d0}}\right)\Bigl[\rho_1(m_1+ 3+ 3\omega_{d0})+ \rho_2(m_2+ 3+ 3 \omega_{d0})\Bigr],\label{DM-present}
\end{align}
%%%%%%%%%%%%%%%%%%%%%%%%%%%%%%%%%%%%%%%%%%%%%%%%%%%%%%%%%%%%%%%%%
where $\omega_{d0}$ is the present value of the EoS in DE\footnote{Let us note that since $\omega_d$ has been considered to be constant, then $\omega_d=  \omega_{d0}$. However, we have used the suffix $0$ in $\omega_d$ to keep the same structure with the other parameters, for instance, $\Omega_{d0}$, $\Omega_{m0}$ and $H_0$. Therefore, in the scenario of interacting DE with constant EoS, $\omega_{d0}$ and $\omega_d$ should be considered to be idential.}. Now,
dividing (\ref{DE-present}) and (\ref{DM-present}) by $\rho_{c0}= 3H_0^2/8\pi G$,
we get the present day density parameters for DE and DM as follows:
\begin{align}
\Omega_{d0} &= -\left(\frac{1}{3 \omega_{d0}}\right)\Bigl[\Omega_1(m_1+ 3)+ \Omega_2(m_2+ 3)\Bigr],\label{DEDP}\\
\Omega_{m0} &=  \left(\frac{1}{3 \omega_{d0}}\right)\Bigl[\Omega_1(m_1+ 3+ 3\omega_{d0})+ \Omega_2(m_2+ 3+ 3\omega_{d0})\Bigr],\label{DMDP}
\end{align}
%%%%%%%%%%%%%%%%%%%%%%%%%%%%%%%%%%%%%%%%%%%%%%%%%%%%%%%%%%%%%%%%%%%%%%%%%%%%%
Now, solving for $\Omega_1$ and $\Omega_2$, we find
\begin{eqnarray}
\Omega_1&=& \frac{(m_2+3+3\omega_{d0})\Omega_{d0}+(m_2+3)\Omega_{m0}}{m_2-m_1},\label{Omega1}\\
\Omega_2&=& \frac{(m_1+3+3\omega_{d0})\Omega_{d0}+(m_1+3)\Omega_{m0}}{m_1-m_2}.\label{Omega2}
\end{eqnarray}
%%%%%%%%%%%%%%%%%%%%%%%%%%%%%%%%%%%%%%%%%%%%%%%%%%%%%%%%%%%%%%%%%%%%%%%%%%%%%%%%%%%%%%%%%%%%%%%%%%

\vspace{0.5cm}

$\bullet$ {\bf When DE is the cosmological constant}\\

Now, we consider the case when DE is the cosmological constant. Thus, in this case, we have $\omega_d= -1$,
hence, $\big(m_1, m_2\big)$ take the forms $\big(0,~-3(1+ \alpha)\big)$.
The total energy density takes the form
\begin{eqnarray}
\rho_t&=& \tilde{\rho}_1+ \tilde{\rho}_2 (1+z)^{3 (1+\alpha)},\label{CC-total}
\end{eqnarray}
%%%%%%%%%%%%%%%%%%%%%%%%%%%%%%%%%%%%%%%%%%%%%%%%%%%%%%%%%%%%%%%%%%%%%%%%%%%%%%%%%%%%%
and the explicit analytic solutions of the energy density for the cosmological constant
and the energy density of the pressureless matter in this interacting picture take the expressions
\begin{eqnarray}
\rho_d&=& \tilde{\rho}_1- \tilde{\rho}_2 \alpha (1+z)^{3(1+\alpha)},\label{CC-density}\\
\rho_m&=& \tilde{\rho}_2 (1+\alpha) (1+z)^{3(1+\alpha)}.\label{DM-density-CC}
\end{eqnarray}
%%%%%%%%%%%%%%%%%%%%%%%%%%%%%%%%%%%%%%%%%%%%%%%%%%%%%%%%%%%%%%%%%%%%
Hence, the Friedmann equation becomes
\begin{eqnarray}
\left(\frac{H}{H_0}\right)^2&=& \tilde{\Omega}_1+ \tilde{\Omega}_2 (1+ z)^{3 (1+ \alpha)},\label{F2}
\end{eqnarray}
%%%%%%%%%%%%%%%%%%%%%%%%%%%%%%%%%%%%%%%%%%%%%%%%%%%%%%%%%%%%%%%%%%%%%%%%%%%%%%%%%%%%%%%%%%%%%
where 
$\tilde{\Omega}_1= \tilde{\rho}_1/\rho_{c0}$, $\tilde{\Omega}_2= \tilde{\rho}_2/\rho_{c0}$,
and the solution has also been found in the context of two non-interacting perfect fluids where one is the Brans-Dicke scalar
field and the other is the perfect fluid (in the form of dust) (Paliathanasis, Tsamparlis, Basilakos, \& Barrow 2016).
Further, we have the following relations:
\begin{eqnarray}
\Omega_{d0}&=& \tilde{\Omega}_1-\alpha \tilde{\Omega}_2 ,\nonumber\\
\Omega_{m0}&=& (1+\alpha)\, \tilde{\Omega}_2.\nonumber
\end{eqnarray}
%%%%%%%%%%%%%%%%%%%%%%%%%%%%%%%%%%%%%%%%%%%%%%%%%%%%%%%%%%%%%%%%%%%%%%%%%%%%%%%%%%%%%%%%%%%%%%%%%%%%%%
Further, we can solve for $\tilde{\Omega}_1$ and $\tilde{\Omega}_2$ as
\begin{eqnarray}
\tilde{\Omega}_2&=& \frac{1}{1+\alpha} \Omega_{m0},\nonumber\\
\tilde{\Omega}_1&=& \Omega_{d0}+ \frac{\alpha}{1+\alpha} \Omega_{m0}.\nonumber
\end{eqnarray}
 %%%%%%%%%%%%%%%%%%%%%%%%%%%%%%%%%%%%%%%%%%%%
It should be noted that, $\Omega_{m0}+\Omega_{d0}= \tilde{\Omega}_1+ \tilde{\Omega}_2= 1$. \\

%\begin{figure}[h]
%%\begin{minipage}{0.45\textwidth}
%\includegraphics[width=5.9cm, height=5cm]{Interacting_Lambda_Omegad_Coupling.eps}
%\caption{This shows 1$\sigma$, 2$\sigma$ confidence levels of the parameters in the plane ($\Omega_d$, $\alpha$) for the data set $SNe+ OHD+BAO$, where $\chi^2= 215.142$, $\Omega_d= 0.669307$, $\alpha= -0.0119593$.}
%\label{Fig:Interacting-Lambda}
%%\end{minipage}
%\end{figure}

\subsection{Asymptotic behavior of the energy densities and the sign of the interaction}
\label{sec:asymptotic}

As the solutions for $\rho_d$, $\rho_m$ include several parameters, so, we shall be very careful, as well as, explicit, while describing the asymptotic behavior of the solutions. First of all, we recall that, the roots $m_1$, $m_2$  should be real in order to realize a feasible state. Hence,
for $m_1$, $m_2$ to be real, we must have $(1-\alpha)^2 \omega_d^2- 4 \alpha \omega_d> 0$. Now, under the low interaction condition, $\alpha^2$ can be neglected, thus, we find that, $(1-2\alpha)\,\omega_d^2- 4 \alpha \omega_d> 0$. As $\omega_d< 0$, we let $\omega_d= -\,d^2$ ($d \in \mathbb{R}$) which gives a restriction as $d^2 (1-2 \alpha)+ 4 \alpha> 0$, which is always true (recall that the interaction is very low, i.e. the coupling parameter is very very small). Thus, if $\omega_d< 0$, then $m_1$, $m_2$ are always real under the low interaction assumption. Now, if $m_1> 0$, then
\begin{eqnarray}
\sqrt{(1-\alpha)^2 \omega_d^2- 4 \alpha \omega_d}\; \;  > (2+ \omega_d-\alpha \omega_d)\nonumber\\
\Longrightarrow 1+\omega_d< 0,~~\mbox{i.e.}~~\omega_d< -1.\nonumber
\end{eqnarray}
%%%%%%%%%%%%%%%%%%%%%%%%%%%%%%%%%%%%%%%%%%%%%%
Similarly, $m_1< 0\Longrightarrow \omega_d> -1$. On the other hand, if $2+ \omega_d- \alpha \omega_d< 0$, we have $\omega_d< -\frac{2}{1-\alpha}< -1$ (in low interaction scheme and if we do not restrict on the coupling parameter, then of course $\alpha< 1$).

In both cases, $m_2< 0$, for $\omega_d< -1$, or $\omega_d> -1$. Hence, we arrive at two different situations: (I) If $\omega_d< -1$, then $m_1> 0$, $m_2< 0$, (II) If $\omega_d> -1$, then $m_1< 0$, $m_2< 0$. Now, we calculate the following expressions in order to analyze the asymptotic behavior of the energy densities.

$$\frac{m_1+ 3}{3 \omega_d}= \frac{1}{2} \left[-(1-\alpha)+ \sqrt{(1-\alpha)^2+ \frac{4 \alpha}{d^2}}\right]> 0$$
$$\frac{m_2+ 3}{3 \omega_d}= \frac{1}{2} \left[-(1-\alpha)+ \sqrt{(1-\alpha)^2+ \frac{4 \alpha}{d^2}}\right]< 0$$

Also, we have the following

\begin{align}
\frac{m_1+ 3+ 3 \omega_d}{3 \omega_d}&=\frac{1}{2} \left[(1+ \alpha)+ \sqrt{(1-\alpha)^2+ \frac{4 \alpha}{d^2}}\right]> 0\nonumber
\end{align}
and
\begin{align}
\frac{m_2+ 3+ 3\omega_d}{3 \omega_d} &= \frac{1}{2} \left[(1+\alpha)+ \sqrt{(1+\alpha)^2+ 4 \alpha\,\left(\frac{1}{d^2}-1\right)}\right]\nonumber\\
&> 0~(\mbox{or}~< 0)~~\mbox{if}~~\omega_d< -1~(\mbox{or}~\omega_d> -1)\nonumber
\end{align}
Now, we discuss the three different stages of the universe as follows:\newline

$\bullet$ For $z \longrightarrow 0$, $\rho_m$, $\rho_d$ are finite, and the total energy density $\rho_t$, as well.

\vspace{0.2cm}

$\bullet$ When $z$ is very large, i.e. $z\longrightarrow \infty$,

\begin{align}
\rho_d &\longrightarrow -\, \frac{\rho_2\,(m_2+ 3)}{3\omega_d} (1+z)^{-m_2};\nonumber\\
\rho_m &\longrightarrow  \frac{\rho_2\,(m_2+ 3+ 3\omega_d)}{3\omega_d} (1+z)^{-m_2};\nonumber\\
\rho_t &\longrightarrow \rho_2 (1+z)^{-m_2}.\nonumber
\end{align}

\vspace{0.2cm}
$\bullet$ When $z \approx -1$, we consider $z= -1+ \epsilon$, where $\epsilon> 0$ is any arbitrary infinitesimal quantity, we have

\begin{align}
\rho_d &\longrightarrow -\, \frac{\rho_1 (m_1+ 3)}{3\omega_d} \epsilon^{-3m_1};\nonumber\\
\rho_m &\longrightarrow \frac{\rho_1 (m_1+ 3+3 \omega_d)}{3\omega_d} \epsilon^{-3m_1};\nonumber\\
\rho_t &\longrightarrow \rho_1 \epsilon^{-m_1}.\nonumber
\end{align}

For convenience, depending on the signs of the quantities, we introduce the following:\newline

$$\frac{m_1+ 3}{3\omega_d}= \beta^2;~~\frac{m_2+ 3}{3\omega_d}= -\, \gamma^2;~~\frac{m_1+ 3+ 3\omega_d}{3\omega_d}= \theta^2$$
$$\frac{m_2+ 3+ 3\omega_d}{3\omega_d}= \delta^2~~(\mbox{if}~\omega_d< -1),~~\mbox{or}~~ = -\,\delta^2~~(\mbox{if}~\omega_d> -1)$$

Thus, based on the nature of the dark energy (that means it is of quintessence or of phantom type), it is possible to express the energy densities of the dark sectors in the following manner.\newline

$\bullet$ If $\omega_d< -1$, then we have

\begin{align}
\rho_d &= -\rho_1 \, \beta^2 (1+z)^{-m_1}+ \rho_2 \, \gamma^2 (1+z)^{-m_2},\nonumber\\
\rho_m &= \rho_1 \theta^2 (1+z)^{-m_1}+ \rho_2 \delta^2 (1+z)^{-m_2}.\nonumber
\end{align}
\vspace{0.2cm}

$\bullet$ If $\omega_d> -1$, we have

\begin{align}
\rho_d &= -\rho_1 \, \beta^2 (1+z)^{-m_1}+ \rho_2 \, \gamma^2 (1+z)^{-m_2},\nonumber\\
\rho_m &= \rho_1 \, \theta^2 (1+z)^{-m_1} - \rho_2 \, \delta^2 (1+z)^{-m_2}.\nonumber
\end{align}

Further, due to the analytic solutions, we can simplify the interaction (\ref{interaction})
into
\begin{eqnarray}
\frac{Q}{H}&=& 3 \alpha H_0^2  (m_1 \Omega_1 a^{m_1}+ m_2 \Omega_2 a^{m_2}).\label{interaction1a}
\end{eqnarray}
%%%%%%%%%%%%%%%%%%%%%%%%%%%%%%%%%%%%%%%%%%%%%%%%%%%%%%%%%%%%%%%%%%%%%%%%%%%%%%%%
Now, from Eq. (\ref{interaction1a}), we can infer about the sign in $Q$ depending on the parameters involved.\\

$\bullet$ If $m_1> m_2> 0$, then $Q> 0$ if $\alpha> 0$; and $Q< 0$ for $\alpha< 0$\\

$\bullet$ Also,  if $0 < m_1 < m_2<$, then $Q> 0$ if $\alpha> 0$; and $Q< 0$ for $\alpha< 0$\\

$\bullet$ If $m_1< m_2< 0$, then $Q> 0$ when $\alpha< 0$; and $Q< 0$ for $\alpha> 0$\\

$\bullet$ Now when $m_1 m_2 < 0$, the sign of $Q$ seems difficult to determine. In fact, in this case, we have an interesting observation which tells that, $Q$ could change its sign during the evolution of the universe. To illustrate the idea, let us take the following simplest example:\\

$\bullet$ \textbf{An Example:} Just for simplicity let us take $m_1= -1$, $m_2= 1$. Therefore, we have $Q= 3 H \alpha H_0^2 \left(-\frac{\Omega_1}{a}+ \Omega_2 a\right)$. Now, the terms in the bracket could be once positive and once negative, which again depends on the sign of $\alpha$ outside the bracket.\\

Hence, generalizing this event (i.e. when $m_1m_2< 0$), the equation $Q= 0$ can have more than one real roots leading to an oscillatory interacting scenarios in which once $Q> 0$, and once $Q< 0$. \\\\

$\bullet$  \textbf{Asymptotic behavior when DE is the cosmological constant}\\

Depending on the coupling parameter $\alpha$, we find the following asymptotic
behavior of the energy densities both at present and late-times.\\

I. When $\alpha> 0$\\

In that case, for $z\longrightarrow 0$, $\rho_d \longrightarrow \tilde{\rho}_1- \alpha \tilde{\rho}_2$ and $\rho_m \longrightarrow \tilde{\rho}_2 (1+ \alpha)$. On the other hand, for $z \longrightarrow \infty$, $\rho_d \longrightarrow -\infty$, $\rho_m \longrightarrow \infty$. Further, in the case for $z \longrightarrow -1$, $\rho_d \longrightarrow \tilde{\rho}_1$, $\rho_m \longrightarrow 0$.\\

II. When $-1< \alpha< 0$\\

For $z\longrightarrow 0$, $\rho_d \longrightarrow \tilde{\rho}_1- \alpha \tilde{\rho}_2 \geq 0$, and $\rho_m \longrightarrow \tilde{\rho}_2 (1+ \alpha)> 0$. When $z \longrightarrow \infty$, $\rho_d \longrightarrow \infty$, and $\rho_m \longrightarrow \infty$. Further, for $z \longrightarrow -1$, $\rho_d \longrightarrow \tilde{\rho}_1$, $\rho_m \longrightarrow 0$.\\

III. When  $\alpha< -1$\\

For $z\longrightarrow 0$, $\rho_d \longrightarrow \tilde{\rho}_1- \alpha \tilde{\rho}_2 \geq 0$, and $\rho_m \longrightarrow \tilde{\rho}_2 (1+ \alpha)< 0$.
For $z \longrightarrow \infty$, $\rho_d \longrightarrow \tilde{\rho}_1$, and $\rho_m \longrightarrow 0$. When $z \longrightarrow -1$, we see $\rho_d \longrightarrow \infty$, $\rho_m \longrightarrow -\infty$.\\

Now, we can rewrite the interaction in this case as
\begin{eqnarray}
\frac{Q}{H}&=& -9 \alpha (1+ \alpha) H_0^2 \tilde{\Omega}_2 a^{-3 (1+ \alpha)},\label{interaction1b}
\end{eqnarray}
%%%%%%%%%%%%%%%%%%%%%%%%%%%%%%%%%%%%%%%%%%%%%%%%%%%%%%%%%%%%%%%%%%%%%%%%%%%%%%%%%%%
which summarizes the following possibilities:\\

$\bullet$ If $\alpha> 0$, then $Q< 0$, i.e. the energy flows from DE to DM.\\

$\bullet$ When $-1 <\alpha < 0$, then $Q> 0$, indicating the flow of energy from DM to DE.\\

$\bullet$ Finally, for $\alpha< -1$, $Q< 0$ leading to the energy flow from DE to DM.\\\\

\section{Interacting dynamics for variable equation of state in DE}
\label{sec-variable-de}

Now, we consider an interacting scenario where  dark energy fluid has a time varying equation of state. This scenario can be considered as a general scenario of interacting dynamics. The possibility of time varying dark energy interacting with pressureless dark matter has been  investigated during last couple of years \cite{He2008, WW2014, NB2014}.
However, the present interaction is different from the others in \cite{He2008, WW2014, NB2014} and it becomes clear when the variable equation of state in DE is considered. So, in the current study along with the constant dark energy equation of state in section \ref{sec-analytic}, we consider the case for dynamical DE.
Let us introduce a generalized equation of state in DE given by \cite{BAZS2009}
\begin{equation}\label{eos-variable}
\omega_d(z)= \omega_{d0} - \omega_{\beta}\, \left(\frac{(1+z)^{-\beta}-1}{\beta}\right),
\end{equation}
where $\omega_{d0}$ is the current value of $\omega_d (z)$, and $\omega_{\beta}$ is the free parameter. We note that the EoS
in (\ref{eos-variable}) recovers three well
known dynamical dark energy equations of state. For $\beta = 1$, we recover the Chevallier-Polarski-Linder (CPL) parametrization \cite{CP2001, Linder2003}
\begin{eqnarray}\label{cpl}
\omega_d(z) = \omega_{d0} + \omega_1\, \left( \frac{z}{1+z} \right ).
\end{eqnarray}
Now, for $\beta= -1$, one finds the linear parametrization \cite{CH1999, Astier2001, WA2002},
\begin{eqnarray}\label{linear}
\omega_d(z) = \omega_{d0} + \omega_{2}\, z,
\end{eqnarray}
while for $\beta \rightarrow 0$, one realizes the logarithmic parametrization in DE \cite{Efstathiou1999}
\begin{eqnarray}\label{log}
\omega_d(z) = \omega_{d0}+ \omega_{3}\, \ln (1+z),
\end{eqnarray}
where $\omega_3$ is the free parameter. Cosequently, for other values of $\beta \neq -1, 0, 1$, one can generate several equations of state.  Thus, one can study the second order differential equation (\ref{ode1}) for the variable EoS in (\ref{eos-variable}) in order to estimate the cosmological parameters with the use of current observational data. On the other hand,
plugging the total energy density $\rho_t= \rho_m + \rho_d= 3 H^2$ into the interaction (\ref{eq-int}) for $\omega_d (z)$ given in eqn. (\ref{eos-variable}), the interaction now becomes
\begin{eqnarray}\label{Q-dynamic}
Q = H^3\, \psi (z),
\end{eqnarray}
where $\psi (z)$ is given by
\begin{align}\label{Q-dynamic1}
\psi(z)= -\, 9\, \alpha \Bigg[ 1+ \left\{\omega_{d0} - \omega_{\beta}\, \left(\frac{(1+z)^{-\beta}-1}{\beta}\right) \right\}\,\Omega_d \Bigg]
\end{align}
and this quantity only determines the sign of $Q$ during the evolution of the Universe (since $H> 0$, for expanding universe). Further,
current value of the interaction rate can be found to be $Q_0= - 9 \alpha H_0^3\, \left(1+ \omega_{d0}\, \Omega_{d0} \right)$ and its sign can be determined once the cosmological parameters are estimated by the current observational data.

Below we shall describe the current observational data to constrain the mentioned  interacting models and from this point of view we aim to find the most successful dynamic EoS given in (\ref{eos-variable}) for the current interacting scenario. For
this purpose we have to solve numerically the second order differential equation
(\ref{ode1}) for $\omega_d(z)$ in (\ref{eos-variable}).

\section{Observational Constraints on the model parameters}
\label{data-analysis}

In order to constrain the free parameters of the models presented in the previous sections, we consider the lastest observational data from Supernovae type Ia (SNIa), Hubble expansion history (OHD), and the
baryonic acoustic oscillation (BAO) distance measurements.
The best fit values, with their corresponding uncertainties, follow from minimizing
the likelihood function
 $\mathcal{L} \varpropto \exp (-\chi^2_{tot}/2)$, where $\chi^2_{tot} = \chi^2_{SN} + \chi^2_{OHD} + \chi^2_{BAO}$.
In the following subsections we briefly describe each data set.

\begin{figure}%[htb]
	\includegraphics[width=8.0cm, height=20.0cm]{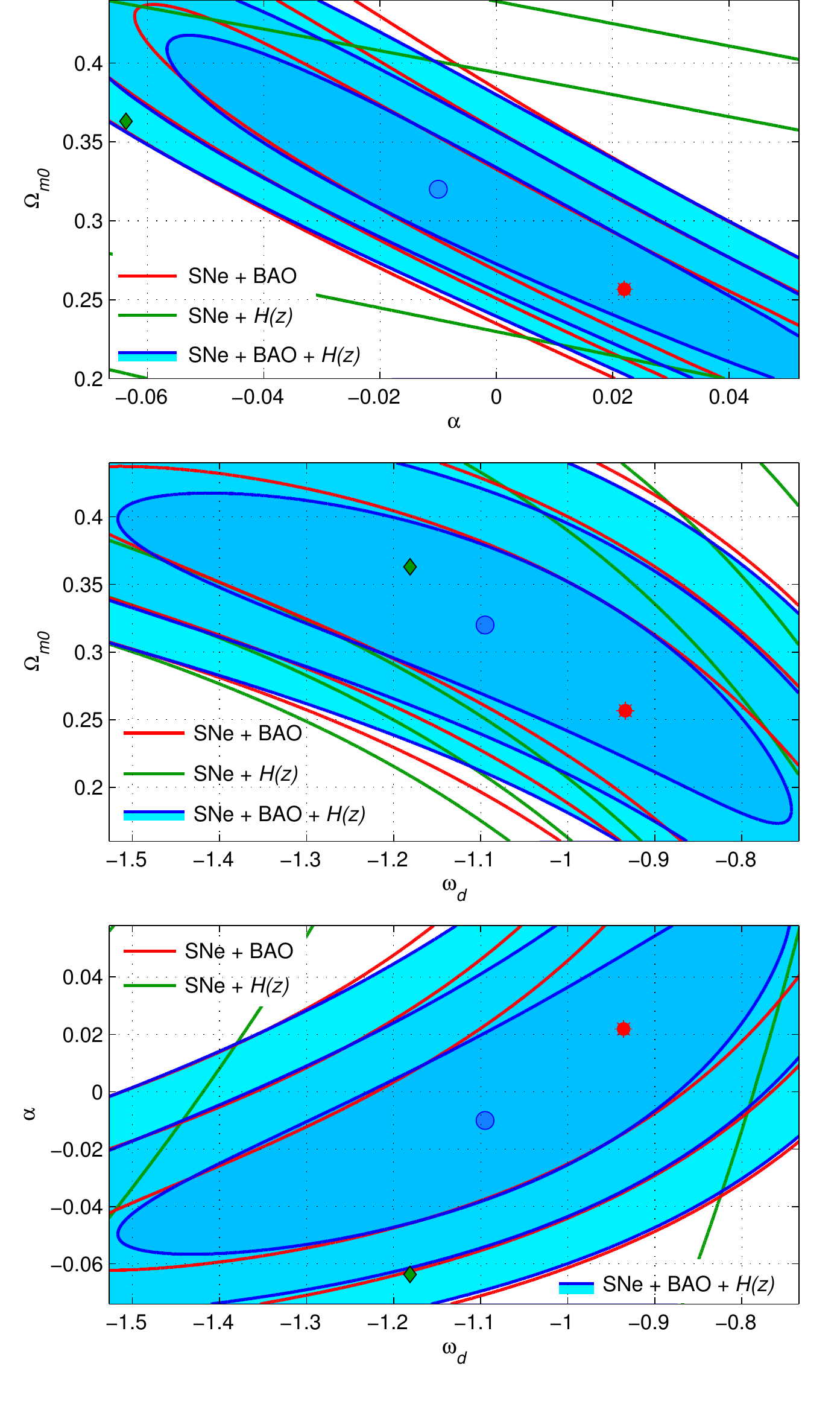}
	\caption{The figure shows the 1$\sigma$, 2$\sigma$, 3$\sigma$ confidence-level contour plots of the model parameters
		with their best fit values for the interacting DE model with constant EoS, $\omega_d \neq -1$ (Eq. (\ref{solution-constant-EoS})),
		using different combinations of the observational data sets as SN$+$ OHD$+$BAO (blue filled contours), SN$+$ BAO (red contours), SNe$+$ OHD (green contours). We note that the contour lines for SN$+$ H($z$) do not appear properly in all the plots. }
	\label{Fig:Constant-EoS}
	%\end{minipage}
\end{figure}

\begin{figure}
	\includegraphics[width=8.9cm, height=7.0cm]{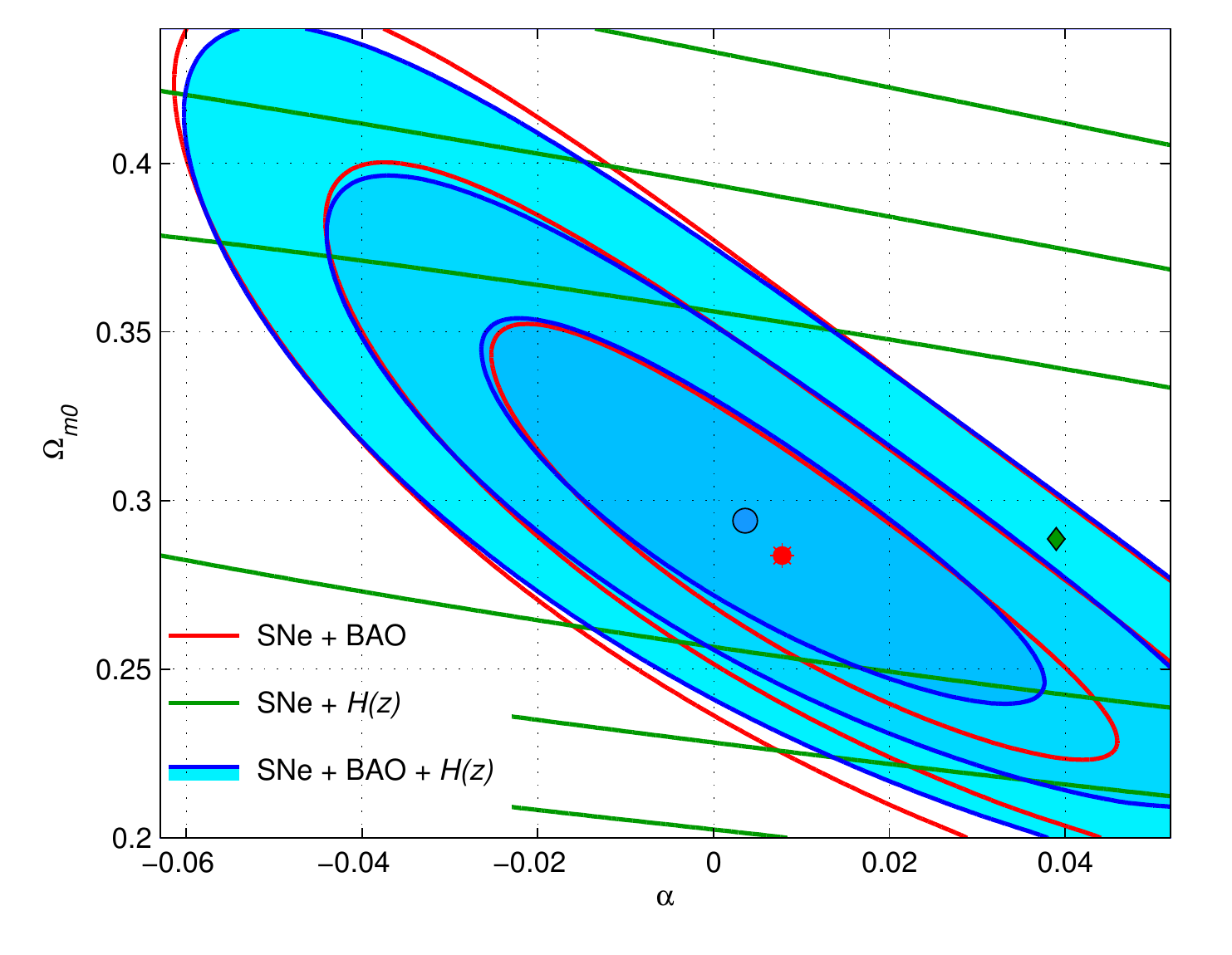}
	\caption{The figure shows the 1$\sigma$, 2$\sigma$, 3$\sigma$ confidence-level contour
		plots of the model parameters with their best fit values for the interacting
		cosmological constant (Eq. (\ref{CC-total})), using different combinations of the
		observational data sets as SN$+$ OHD$+$BAO (blue filled contours), SN$+$ BAO (red
		contours), SNe$+$ OHD (green contours).} \label{Fig:Interacting-Lambda}
	%\end{minipage}
\end{figure}

\subsection{Supernovae Type Ia data}

Data from Supernovae was the first indication for accelerating universe, and it is very useful to test the cosmological models. In the present analysis we use
 580 Supernovae data points from Union 2.1 available in \cite{Suzuki2012}
 in the redshift interval $0 \leq z \leq 1.41$. Now, for any Supernova, situated at a
redshift $z$, the distance modulus $\mu$ is given by
\begin{eqnarray}
\mu (z)&=& 5 \log_{10} \left(\frac{d_L(z)}{1 \mbox{Mpc}}\right)+ 25,\label{mu}
\end{eqnarray}
%%%%%%%%%%%%%%%%%%%%%%%%%%%%%%%%%%%%%%%%%%%%%%%%%%%%%%%%%%%%%%%%%%%%%%%%%%%%%%%%%%%%%%%%%%%%%%%%%%
where $d_L (z)= c (1+z) \int _0 ^ z \frac{dz^\prime}{H (z^\prime)}$ is the luminosity distance.
The $\chi^2$ function is evaluated as
\begin{equation}
\chi^2_{SN}(\theta_1,\dots)= \min\limits_{H_0}\sum_{i,j=1}^{580}
 \Delta\mu_i\big(C_{SN}^{-1}\big)_{ij} \Delta\mu_j.\label{chi2a}
\end{equation}
 where $\Delta\mu_i=\mu_{th}(z_i,\theta_1,\dots)-\mu_{obs}(z_i)$, $\theta_k$, $C_{SN}$ are respectively the discrepancy between theory and observations, model
 parameters to be fitted,  and the $580\times580$ covariance matrix \cite{Suzuki2012}.
We use the function (\ref{chi2a}), marginalized over the Hubble constant $H_0$.

\subsection{Hubble data}

Hubble parameter measurements are also useful to constrain the parameters in dark energy
models. Two independent methods have been widely used to measure the Hubble parameter
values at different redshifts. One is the differential age of some galaxy, and the other
one is the radial BAO size methods.
Here we use 30 OHD points for our
    analysis from Refs. \cite{Simon2005, Stern2010, Zhang2014, Moresco2012, Moresco2015, Moresco2016}, obtained with the
    differential age method. To constrain the model parameters, we apply the $\chi^2$ function by the
    following method
\begin{eqnarray}
\chi^2_{OHD}&=& \sum_{i=1}^{30} \left[\frac{H_{obs}(z_i)-H_{th}(z_i,
    \theta_j)}{\sigma_i^2}\right],\label{chi-OHD}
\end{eqnarray}
%%%%%%%%%%%%%%%%%%%%%%%%%%%%%%%%%%%%%%%%%%%%%%%%%%%%%%%%%%%%%%%%%%%%%%%%%%%%%%%%%%
where $\theta_j$ stands for different model parameters to be constrained.

\subsection{Baryon Acoustic Oscillation data}

To analyze cosmological models, baryon acoustic oscillations (BAO) data give another useful test. Here, we use BAO data, i.e. $\frac{d_A(z_{\ast})}{D_V (Z_{BAO})}$ from the references \cite{Blake2011, Percival2010, Beutler2011, Jarosik2011}, where
$z_{\ast}$ is the radiation-matter decoupling time given by $z_{\ast} \approx 1091$, $d_A$ is the co-moving angular diameter distance given by $d_A= \int_0 ^z \frac{dz}{H(z)}$, and, $D_V= \left(d_A (z)^2 \frac{z}{H(z)}\right)^{1/3}$ is the dilation scale \cite{Eisenstein2005}. In table \ref{bao-table}, we present the observational BAO data. We use the following procedure as in \cite{Giostri2012} to calculate the $\chi^2$ minimization for BAO data, i.e.
\begin{eqnarray}
\chi^2_{BAO}&=& X^T_{BAO} C^{-1}_{BAO} X_{BAO},\label{chi2-bao}
\end{eqnarray}
%%%%%%%%%%%%%%%%%%%%%%%%%%%%%%%%%%%%%%%%%%%%%%%%%%%%%%%%%%%%%%%%%%%%%%%%%%%%%%%%%%%%%%%%%%%%%%%%%%%%%%%%%%%%%%%%%%%%%%%%
where

\[
X_{BAO}
=
\begin{bmatrix}
    \frac{d_A(z_{\ast})}{D_V(0.106)}-30.95 \\
    \frac{d_A(z_{\ast})}{D_V(0.2)}-17.55 \\
    \frac{d_A(z_{\ast})}{D_V(0.35)}-10.11\\
    \frac{d_A(z_{\ast})}{D_V(0.44)}-8.44\\
        \frac{d_A(z_{\ast})}{D_V(0.6)}-6.69\\
        \frac{d_A(z_{\ast})}{D_V(0.73)}-5.45
\end{bmatrix}
\]
and the inverse covariance matrix $C^{-1}$ can be found in (Giostri et al. 2012).

\begin{table*}
%\begin{center}
\caption{Values of $\frac{d_A (z_{\ast})}{D_V (z_{BAO})}$ for different values of $z_{BAO}$.}
%\begin{adjustbox}{angle=90}
\begin{tabular}{ccccccc}
\hline
%\hline
$z_{BAO}$~~ & 0.106~~& 0.2~~ & 0.35~~ & 0.44~~ & 0.6~~ & 0.73\\
\hline
\hline
$\frac{d_A (z_{\ast})}{D_V (z_{BAO})}$~~ & 30.95 $\pm$ 1.46~~ & 17.55 $\pm$ 0.60~~ & 10.11 $\pm$ 0.37~~ & 8.44 $\pm$ 0.67~~ & 6.69 $\pm$ 0.33~~ & 5.45 $\pm$ 0.31\\
\hline
%\end{center}
\end{tabular}
%\caption{Values of $\frac{d_A (z_{\ast})}{D_V (z_{BAO})}$ for different values of $z_{BAO}$.}
\label{bao-table}
%\end{adjustbox}
\end{table*}

\section{Results of the joint analysis}
\label{results}

In this section we shall present the main observational results extracted from the scenarios of interacting dark energy when dark energy equation of state is either constant or evolving with the cosmic time. In case of constant EoS in DE we have considered two separate cosmological models one is the interacting DE model with $\omega_d \neq -1$, and interacting cosmological constant where for both the models the background evolution equations are analytically solved. On the other hand, for the variable EoS in DE we have considered a general EoS given in (\ref{eos-variable}) and solved the second order differential equation
(\ref{ode1}) numerically. In the following we shall describe the observational constraints on the model parameters for the interacting dark energy models.

\subsection{Constant equation of state in DE}

We fit both the interacting models with constant EoS using the latest observational data
from Supernovae Type Ia, observed Hubble parameter data, and the baryon acoustic oscillations distance measurements. We note that for all interacting models, we have
marginalized $\chi^2_{tot}$ for all values of $H_0$. Let us describe the observational
constraints and the main cosmological features of the models separately as follows.\\

 $\bullet$ For interacting dark energy, where $\omega_d \neq - 1$,
$\chi^2_{min} (\Omega_{m0}, \alpha, \omega_{d0})= 562.39$, $\omega_{d0}=
-1.095_{-0.421}^{+0.352}$, $\Omega_{m0}= 0.320_{-0.147}^{+0.093}$, $\alpha=
-0.010_{-0.046}^{+0.01}$. Here acceptable values of $\alpha$, i.e its priors are limited
to $\alpha<0$, because for positive $\alpha$ we have solutions with negative dark energy
density $\rho_d$ given in eqn. (\ref{DE-constantA}) at early stages of evolution. They may be considered as singular solutions, though the total density $\rho=\rho_m+\rho_d$
remains positive and the scale factor $a(t)$ behaves regularly.
 Also, the reduced $\chi^2= \chi^2_{min}/d.o.f= 0.917$ ($d.o.f=$ degrees of freedom $=$
$N-P$, where $N$ is total number of data points; $P$ is total number of independent
model parameters). In figure Fig.~\ref{Fig:Constant-EoS} we have shown the contour plots
at 1$\sigma$, 2$\sigma$ and 3$\sigma$ confidence levels with different combinations of
the observational data sets, namely, SN$+$BAO, SN$+$OHD and SN$+$OHD$+$BAO.

Therefore, from the analysis of this model, it is seen that this interaction  can describe the phantom nature of the universe.
Further, concerning the interaction $Q$ in eqn. (\ref{interaction1a}), using the best fit values of the cosmological parameters,
one can see that $m_1 > 0$, but $m_2 < 0$, which indicates that during the evolution of the universe, $Q$ should change its sign, but however,
 the observational data tells that for this scenario, $Q > 0$, for any $z \geq 0$, that means the energy flows from DE to DM.  \\

$\bullet$ For interacting $\Lambda$ with $\omega_d = - 1$, we found
$\chi^2_{min} (\Omega_{m0},\alpha)= 562.55$, hence, reduced $\chi^2= \chi^2_{min}/d.o.f=
0.919$. It is achieved for $\Omega_{m0}= 0.294_{-0.054}^{+0.060}$, $\alpha=
0.0035_{-0.028}^{+0.002}$.  In Fig.~\ref{Fig:Interacting-Lambda} we have shown the
contour plot for $(\Omega_{m0}, \alpha)$ at 1$\sigma$, 2$\sigma$ and 3$\sigma$
confidence levels for three different combinations of the observational data sets,
namely SN$+$BAO, SN$+$OHD and SN$+$OHD$+$BAO.
  The positive optimal value of $\alpha$ includes singular
behavior of DE density $\rho_d$. In this case 
from eqn. (\ref{interaction1b}) one finds that $Q < 0$, that means, there is an energy
flow from DM to DE. If we forbid solutions with $\alpha> 0$, the $\Lambda$CDM case
($\alpha=0$) will be more preferable.

\begin{table*}
    \begin{center}
\caption{The table summarizes the best fit values of the model parameters with their
errors bars at $1\sigma$ confidence level using the observational data SN$+$OHD$+$ BAO.
$M_1$ is the interacting DE with constant EoS ($\omega_d \neq -1$);
$M_2$ is the interacting cosmological constant; $M_3$ is the interacting DE with CPL parametrization (\ref{cpl}). }
        \begin{tabular}{|c|c|c|c|c|c|}
            \hline
            \hline
            Models &  $\chi^2_{min} (\theta_i)$ & Model parameters & $\chi^2_{min}/d.o.f$ & $AIC$ &  $|\Delta AIC_{1,i}|$\\
            \hline
            $\Lambda$CDM   &  562.58 & $\Omega_{m0}= 0.299_{-0.033}^{+0.035}$ & 0.915 &  564.58 & 0 \rule{0pt}{1.2em}     \\
            $M_1$ & 562.39  & $\omega_d= -1.095_{-0.421}^{+0.352}$, $\Omega_{m0}= 0.320_{-0.147}^{+0.093}$, $\alpha= -0.010_{-0.046}^{+0.01}$ & 0.917  &568.39  &  3.81\rule{0pt}{1.2em} \\
            $M_2$   &  562.55 &  $\Omega_{m0}= 0.294_{-0.054}^{+0.060}$, $\alpha= 0.0035_{-0.028}^{+0.002}$ & 0.916 & 566.55 & 1.97  \rule{0pt}{1.2em}     \\
            $M_3$ & 562.21  & $\omega_{d0}= -1.018_{-0.46}^{+0.465}$, $\Omega_{m0}= 0.378_{-0.066}^{+0.074}$, $\alpha=-0.026_{-0.031}^{+0.026}$, & 0.919  &570.21  &  5.63 \rule{0pt}{1.2em} \\
            &   &  $\omega_1= -1.74_{-4.22}^{+2.76}$ &  &   & \rule{0pt}{1.2em}  \\
            \hline \hline
        \end{tabular}
        \label{table:dataAnalysis}
    \end{center}
\end{table*}

\begin{figure*}
    \includegraphics[width=17.5cm, height=12.0cm]{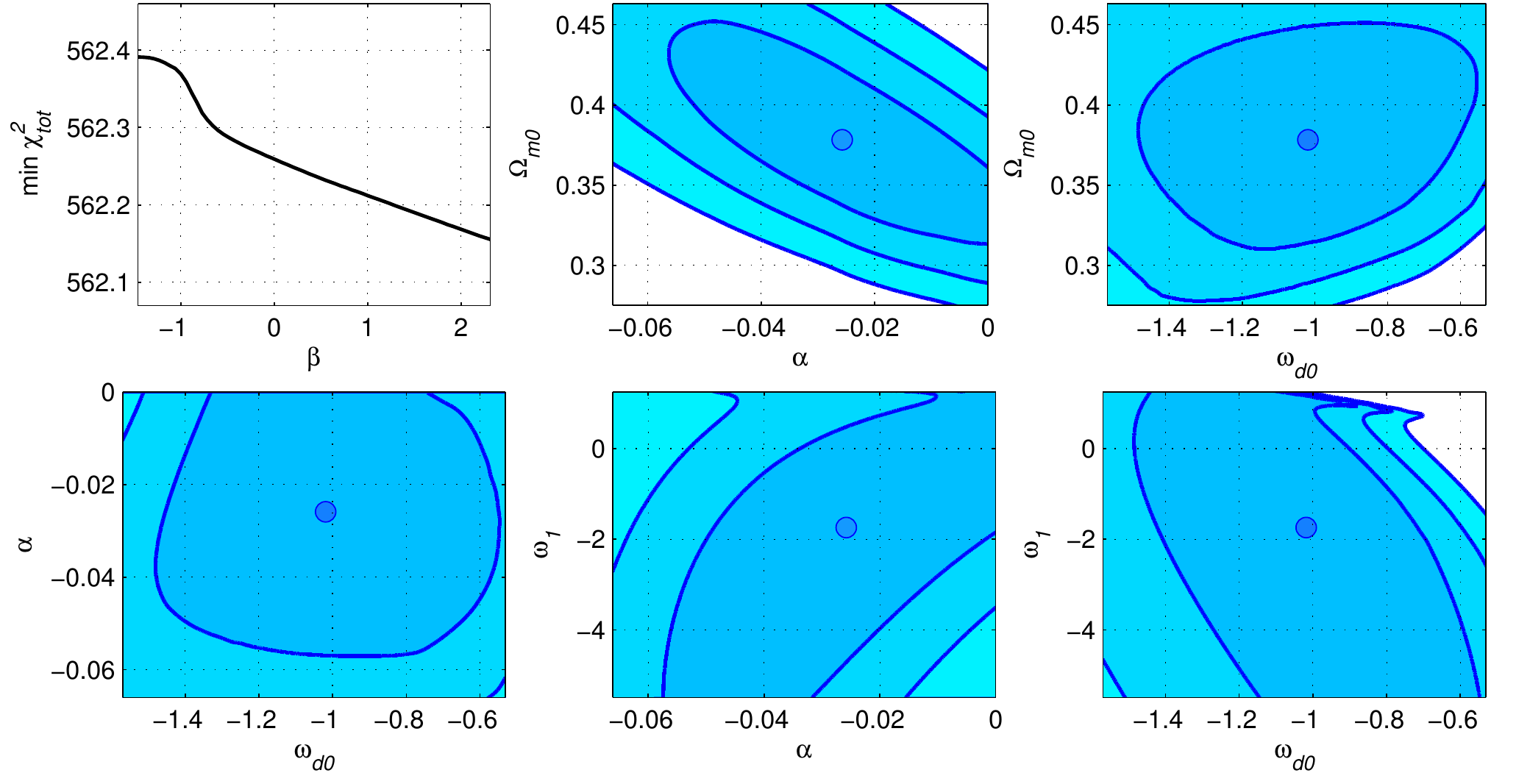}
    \caption{In the top-left panel we show the dependence of min\,$\chi^2_{tot}$ on the free parameter `$\beta$' of the general EoS given in (\ref{eos-variable}) using the  observational data SN$+$OHD$+$BAO. It shows that $\beta= 1$, i.e. the interacting model with CPL parametrization (eqn. (\ref{cpl})) is the viable interacting model with variable EoS in DE in compared to the others. The other panels represent the contour plots at 1$\sigma$, 2$\sigma$ confidence levels for various quatities of the interacting DE model with CPL parametrization (eqn. (\ref{cpl}), i.e. $\beta=1$) using the same observational data SN$+$OHD$+$BAO.} \label{Fig:3}
\end{figure*}

\subsection{Variable equation of state in DE}

We consider the scenario of  interacting dark energy with variable EoS in the form  (\ref{eos-variable}) 
(Barboza, Alcaniz, Zhu \& Silva 2009). This model has 3 free parameters  $\omega_{d0}$, $\omega_{\beta}$, $\beta$ in addition to $\Omega_{m0}$, and $\alpha$ 
(since for all models we marginalize over $H_0$).
This number of model
parameters $P=5$ is too large,
if we take into account information criteria, described in the next section.
So we have to search the most suitable value of $\beta$ in EoS  (\ref{eos-variable}), in
other words, the best choice among the most popular scenarios (\ref{cpl}),
(\ref{linear}), (\ref{log}).  These scenarios are particular cases of the general EoS
(\ref{eos-variable}), if $\beta=1,$ $-1$, 0 correspondingly; in all these cases we
diminish $P$ to $P=4$. To investigate the generalized EoS (\ref{eos-variable}), we calculate how the minimum 
% $\min\limits_{\Omega_{m0},\alpha,H_0,\omega_{d0},\omega_{\beta}}\chi^2_{tot}$ 
 $\min\limits_{\Omega_{m0},\alpha,\omega_{d0},\omega_{\beta}}\chi^2_{tot}$  
(over all other parameters) depends on $\beta$. This dependence is shown in the top-left panel of
figure~\ref{Fig:3}. We see that the best values of $\min\chi^2_{tot}$ correspond to
large positive $\beta$ (moreover, negative values of $\beta$ lead to additional
singularities), so we should choose the CPL parametrization (\ref{cpl}) (Chevallier \&
Polarski 2001; Linder 2003), corresponding to $\beta=1$, as the most successful scenario
of  interacting DE with variable EoS. For this scenario we draw the contour plots for various quantities for the observational data SN$+$OHD$+$BAO in other panels of Fig.~\ref{Fig:3}. The correspondent estimations
of $\min\chi^2_{tot}$ and model parameters are presented in table
\ref{table:dataAnalysis}. As we observe, the interacting CPL model shows that the
current dark energy EoS has a slight phantom nature (as $\omega_{d0}= -1.018< -1$), which is although very close to the $\Lambda$-cosmology. On the other hand, using the observational data into eqns. (\ref{Q-dynamic}) and (\ref{Q-dynamic1}), it is easy to see that at present time, i.e. at $z= 0$, $Q > 0$. 

\subsection{Model selection}

In order to test the quality of fit of the present models by information criteria, we apply the \textit{Akaike Information Criterion} (AIC) \cite{Akaike1974}. The AIC is defined as
\begin{eqnarray}
&& AIC  = -2 \ln \mathcal{L}+ 2 k = \chi^2_{min} + 2 k,\label{aic}
\end{eqnarray}
where $\mathcal{L}= \exp\left(-\chi_{min}^2/2\right)$ is the maximum likelihood function, $k$ is
the number of model parameters and $N$ denotes the
total number of data points used in the statistical analysis. However, to quantify any cosmological
model, a reference model is needed and $\Lambda$CDM is of course the best choice for that.
Now, for any concerned model $M$, other than the reference model (denoted by $R$), from the difference
$\Delta AIC_{M,R}= AIC_{M}- AIC_{R}$,
we arrive at the following conclusions as described in \cite{aic-bic}: (i) If $\Delta AIC_{M, R} \leq 2$, then the concerned model has
substantial support with respect to the reference model (i.e. it has evidence to be a good cosmological model),
(ii) $4 \leq \Delta AIC_{M, R} \leq 7$ indicates less support with respect to the reference model,
and finally, (iii) $\Delta AIC_{M, R} \geq 10$ means that the model has no support, in fact it has no use in principle.
For simplicity we denote the interacting dark energy model where $\omega_d \neq -1$ by $M_1$; interacting cosmological constant by $M_2$ and interacting CPL as $M_3$.
Therefore, from our analysis (see
table \ref{table:dataAnalysis}), we find that $\Delta AIC_{M_1,R}= 1.97 < 2$
$\Delta AIC_{M_2,R} = 3.81< 4$, and $\Delta AIC_{M_3, R}= 5.63 > 4$. Therefore, from AIC it is evident that $M_2$ is the most favored interacting model in compared to the reference model $\Lambda$CDM. However, the model $M_1$ is also supported by AIC analysis, while the interacting CPL has less support in compared to the reference model.

\section{Summary and Discussions}
\label{discu}

In this work, considering a spatially flat FLRW universe, we have
described the dynamics of two barotropic cosmic fluids, namely, pressureless DM and DE which are interacting with each other. The interaction is nongravitational and this
essentially indicates an exchange of
energy and/or momentum between the interacting components, and hence determines the dominant
character among the components in the cosmic sector during the evolution of the universe. We specify the nongravitational interaction given in equations (\ref{interaction}) or (\ref{eq-int}) characterized by a single coupling parameter `$\alpha$', and the EoS in DE has been considered to be either constant or variable with the cosmic evolution. 

We show that if the dark energy equation of state is supposed to be constant with the evolution of the universe,
then the evolution equations for CDM and DE, and hence the Hubble parameter can be completely
solved as a function of the scale factor/redshift, where the standard
evolution equations  $\rho_m \propto a^{-3}$, $\rho_d \propto a^{-3(1+\omega_d)}$, are recovered under the no-interaction limit, that means for $\alpha \rightarrow 0$.
We present two separate sets of solutions when $\omega_d \neq -1$, and $\omega_d= -1$ (i.e. cosmological constant). Additionally,
we present an asymptotic analysis of the solutions in order to evaluate their behavior in the extreme limit, and analyzed the nature of the interaction $Q$ which predicts a possible sign change during its evolution.

All the models (for constant and variable EoS in DE) have been constrained with the latest observational data from (i) Supernovae Type Ia, (ii) observed Hubble parameters, and (iii) Baryon acoustic oscillations distance measurements. The results have been summarized in table \ref{table:dataAnalysis} which shows that for the interacting model with $\omega_d \neq -1$, the observational data slightly favor $\omega_d< -1$, that means the dark energy EoS crosses the phantom divide line `$-1$'. The crossing of phantom divide line is not new in the literature, see for instance \cite{Planck2014, Planck2015, Rest2014, Xia2013, Cheng-Huang, Shafer}.  Consequently, the observational constraints further show that $Q$ might have transient nature, that means during the universe evolution it may change its sign, but for $z \geq 0$, $Q >0$. On the other hand, for interacting cosmological constant (i.e. $\omega_d= -1$) the observational data prefer a nonzero coupling parameter which is positive and very close to zero and consequently from (\ref{interaction1b}), this model gives $Q < 0$. 

Now, in case of interacting dynamical DE, we have considered a general equation of state classified by a sole parameter `$\beta$' (see eqn. (\ref{eos-variable})) which recovers three known and most used DE parametrizations, namely for $\beta = 1$ one has CPL parametrization \cite{CP2001, Linder2003}; $\beta = -1$ gives linear parametrization \cite{CH1999, Astier2001, WA2002}; $\beta \rightarrow 0$ provides the logarithmic parametrization \cite{Efstathiou1999} and consequently, one can find several new DE parametrizations with other values of $\beta \neq -1, 0, 1$. Now, from the $\chi^2$ analysis, using the observational data SN$+$OHD$+$ BAO, we found that the present interaction prefers CPL parametrization from the other models (see the top left panel in Fig. \ref{Fig:3}) and hence we fixed our aim to interacting DE with CPL parametrization and analyzed the cosmological scenario. The constraints on the free model parameters have been summarized in table \ref{table:dataAnalysis}. The contour plots of various quantities in interacting CPL model have been shown in other panels of figure \ref{Fig:3} for the joint analysis SN$+$OHD$+$ BAO. The analysis shows that the observational data allows a nonzero coupling parameter which is negative but close to zero, and it also favors a slight phantom nature of DE similar to the case of interacting DE with constant $\omega_d \neq -1$. Additionally, at present epoch, i.e. at $z= 0$, the interaction is positive ($Q > 0$). 

Moreover, based on the \textit{Akaike Information Criterion} ($AIC$) \cite{Akaike1974}, we see that the interacting models with constant EoS are favored than the interacting dynamical dark energy with CPL parametrization.
In fact, among the three interacting models, the $AIC$ information criterion selects interacting model with $\omega_d= -1$ to be the best fitted cosmological model with the observational data.

Summarizing, for the current interaction model between DE and DM the observational data depict a nonzero but very small interaction between the dark sectors and on the other hand, although all the interaction models are qualitatively very close to the $\Lambda$CDM cosmology, but the models with constant EoS in DE are favored by the AIC criterion than the model with varibale EoS in DE.

\section*{Acknowledgments}

The authors thank the referee for some effective and illuminating comments to improve the work. SP acknowledges Science and Engineering Research Board (SERB), Govt. of India, for
awarding National Post-Doctoral Fellowship (File No: PDF/2015/000640) and the Department of Mathematics, Jadavpur University where a part of the work was completed.
SC thanks IUCAA, Pune, India, for their warm hospitality while working on this project. Also, SP thanks Dr. A. Paliathanasis and Dr. N. Tamanini for helpful discussions.

%%%%%%%%%%%%%%%%%%%%%%%%%%%%%%%%%%%%%%%%%%%%%%%%%%%%%%%%%%%%%%%%%%%%%%%%%%%%%%%%%%%%%%%%%%%%%%%%%%%%%%%%%%%%

\end{document}